\documentclass[twoside,twocolumn,9pt]{article}
\usepackage{extsizes}
\usepackage[super,sort&compress,comma]{natbib} 
\usepackage[version=3]{mhchem}
\usepackage[left=1.5cm, right=1.5cm, top=1.785cm, bottom=2.0cm]{geometry}
\usepackage{balance}
\usepackage{times,mathptmx}
\usepackage{sectsty}
\usepackage{graphicx} 
\usepackage{lastpage}
\usepackage{amssymb}
\usepackage[format=plain,justification=justified,singlelinecheck=false,font={stretch=1.125,small,sf},labelfont=bf,labelsep=space]{caption}
\usepackage{float}
\usepackage{fancyhdr}
\usepackage{fnpos}
\usepackage[english]{babel}
\addto{\captionsenglish}{%
  \renewcommand{\refname}{Notes and references}
}
\usepackage{array}
\usepackage{droidsans}
\usepackage{charter}
\usepackage[T1]{fontenc}
\usepackage[usenames,dvipsnames]{xcolor}
\usepackage{setspace}
\usepackage[compact]{titlesec}
\usepackage{hyperref}
\usepackage{cancel}
\usepackage[normalem]{ulem}
\usepackage{tikz}
\usepackage{soul,xcolor}

\usepackage{epstopdf}

\definecolor{cream}{RGB}{222,217,201}

\begin{document}

\pagestyle{fancy}
\thispagestyle{plain}
\fancypagestyle{plain}{

\renewcommand{\headrulewidth}{0pt}
}

\makeFNbottom
\makeatletter
\renewcommand\LARGE{\@setfontsize\LARGE{15pt}{17}}
\renewcommand\Large{\@setfontsize\Large{12pt}{14}}
\renewcommand\large{\@setfontsize\large{10pt}{12}}
\renewcommand\footnotesize{\@setfontsize\footnotesize{7pt}{10}}
\renewcommand\scriptsize{\@setfontsize\scriptsize{7pt}{7}}
\newcommand\redsout{\bgroup\markoverwith{\textcolor{red}{\rule[0.5ex]{2pt}{0.4pt}}}\ULon}
\makeatother

\renewcommand{\thefootnote}{\fnsymbol{footnote}}
\renewcommand\footnoterule{\vspace*{1pt}%
\color{cream}\hrule width 3.5in height 0.4pt \color{black} \vspace*{5pt}} 
\setcounter{secnumdepth}{5}

\makeatletter 
\renewcommand\@biblabel[1]{#1}            
\renewcommand\@makefntext[1]%
{\noindent\makebox[0pt][r]{\@thefnmark\,}#1}
\makeatother 
\renewcommand{\figurename}{\small{Fig.}~}
\sectionfont{\sffamily\Large}
\subsectionfont{\normalsize}
\subsubsectionfont{\bf}
\setstretch{1.125} 
\setlength{\skip\footins}{0.8cm}
\setlength{\footnotesep}{0.25cm}
\setlength{\jot}{10pt}
\titlespacing*{\section}{0pt}{4pt}{4pt}
\titlespacing*{\subsection}{0pt}{15pt}{1pt}

\fancyfoot{}
\fancyhead{}
\renewcommand{\headrulewidth}{0pt} 
\renewcommand{\footrulewidth}{0pt}
\setlength{\arrayrulewidth}{1pt}
\setlength{\columnsep}{6.5mm}
\setlength\bibsep{1pt}

\makeatletter 
\newlength{\figrulesep} 
\setlength{\figrulesep}{0.5\textfloatsep} 

\newcommand{\topfigrule}{\vspace*{-1pt}%
\noindent{\color{cream}\rule[-\figrulesep]{\columnwidth}{1.5pt}} }

\newcommand{\botfigrule}{\vspace*{-2pt}%
\noindent{\color{cream}\rule[\figrulesep]{\columnwidth}{1.5pt}} }

\newcommand{\dblfigrule}{\vspace*{-1pt}%
\noindent{\color{cream}\rule[-\figrulesep]{\textwidth}{1.5pt}} }

\newcommand{\hcancel}[1]{%
    \tikz[baseline=(tocancel.base)]{
        \node[inner sep=0pt,outer sep=0pt] (tocancel) {#1};
        \draw[red] (tocancel.south west) -- (tocancel.north east);
    }%
}%
\newcommand{\msout}[1]{\text{\sout{\ensuremath{#1}}}}
\setstcolor{red}

\makeatother

\twocolumn[
  \begin{@twocolumnfalse}
\vspace{3cm}
\sffamily
\begin{tabular}{m{4.5cm} p{13.5cm} }

& 
\noindent\LARGE{\textbf{Length-scales of Dynamic Heterogeneity in a Driven Binary Colloid}} \\
& \vspace{0.3cm} \\

& \noindent\large{ S. Dutta $^{{a,b},{\ast}}$ and J. Chakrabarti $^{{a}}$} \\

\end{tabular}

 \end{@twocolumnfalse} \vspace{0.6cm}
]

\renewcommand*\rmdefault{bch}\normalfont\upshape
\rmfamily
\section*{}
\vspace{-1cm}


\footnotetext{$^{a}$ S. N. Bose National Centre for Basic Sciences, Blcok-JD, Sector-III, Salt Lake, Kolkata 700 106}

\footnotetext{$^{b}$ Present Address: The Institute of Mathematical Sciences, 4th Cross Road, CIT Campus, Taramani, Chennai 600113}
\footnotetext{$\ast~$ Correspondence to: sumand@imsc.res.in}



\sffamily{\textbf{ Here we study characteristic length scales in an aqueous suspension of symmetric oppositely charged colloid subject to a uniform electric field by Brownian Dynamics simulations. We consider a sufficiently strong electric field where the like charges in the system form macroscopic lanes. We construct spatial correlation functions characterizing structural order and that of particles of different mobilities in-plane transverse to the electric field at a given time. We call these functions as equal time density correlation function (ETDCF). The ETDCF between particles of different charges, irrespective of mobilities, are called structural ETDCFs, while those between particles of different mobilities are called the dynamic ETDCF. We extract the characteristic length of correlation by fitting the envelopes of the ETDCFs to exponential dependence. We find that structural ETDCF and the dynamical-ETDCFs of the slow particles increase with time. This suggests that the slow particles undergo microphase separation in the background of the fast particles which drive the structural pattern in the plane transverse to the lanes. The ETDCFs can be measured for colloidal systems directly following particle motion by video-microscopy and may be useful to understand patterns out of equilibrium. }}\\

\rmfamily 


\section{Introduction}

Dynamic heterogeneity describes a situation where particles of different mobilities coexist in a system, resulting in distinct relaxation of local spatial order\cite{dh1}. Such spatio-temporal inhomogeneity in dynamics is seen in a broad range of systems, spanning from super-cooled liquids \cite{dh1} to bio-molecular suspensions \cite{wang}. Recently there has been attention on the spatial distributions of particles with different mobilities in exploring dynamic heterogeneity \cite{dutta3, tanaka}. It is conceptually tailored decades back \cite{donati} to understand the onset of slowing down and amorphous order accompanied by a growth in dynamic length-scale when there are only subtle changes in the geometric structure \cite{tanaka,donati}. However, the existence of dynamical length-scale is yet debated in the literature \cite{dh1, tanaka, smk1}. This calls for establishing the existence of such length-scale in a system where experiments can be unambiguously performed by following the particle motions. Colloidal suspensions are ideal to this end \cite{soft1,rev1a,rev1b,rev2a,rev2b}.

 When an aqueous suspension of binary colloid with opposite charges is driven by a time-independent and uniform electric field, the system evolves to micro-phase segregation with like charged particles proliferated along the electric field forming lanes \cite{exp1}. The laning transition in colloids has been observed experimentally\cite{exp1,exp2,exp3,plasma}, and extensively studied in numerous simulations and theoretically \cite{lane1,lane2,lane3,lane3a,netz,lane5,lane9,lane9a,lane10,lane11,lane11a,lane9b,lane13} as a prototype of pattern formation away from equilibrium\cite{dutta3,lane10,lane13}. These studies suggest heterogeneity in the dynamics of particles where particles in the lanes are dynamically different from those not in the lanes\cite{dutta3,lane13,exp2}. In recent studies\cite{lane10,lane13}, we have considered structural patterns in the plane perpendicular to the electric field. Our studies show non-Gaussian behaviour in the probability distributions of particle displacements and anomalous structural relaxation in this plane\cite{lane10}. Such behaviour has been phenomenologically ascribed to heterogeneity in the single-particle diffusion \cite{wang} and has been observed in our studies also \cite{lane10,lane13}. However, there has not been attempt so far to examine the spatial distribution of particles with different diffusivity which we study here. Our objective is two-fold: to characterize spatial arrangement of particles of different mobilities and to understand the structural development in the system through such arrangements. We expect to throw useful lights in understanding kinetics of pattern formation out of equilibrium \cite{kin1,kin2, kin3} and dynamically arrested states\cite{kin3a,kin3b,kin4,kin5} which are relevant in a variety of condensed matter systems\cite{appl}. 

We take an equimolar symmetric binary mixture of oppositely charged colloidal suspension in water, subject to a constant unidirectional electric field as in our earlier studies\cite{lane13}. We perform Brownian Dynamics (BD) simulation\cite{bd1,bd2,hansen-mcdonald} on the system repeating over a number of noise realizations. In these simulations, the solvent is not explicitly considered, but taken into account via the medium viscosity and random forces on the colloidal, following the standard approach of describing colloidal motion \cite{bd1,bd2,hansen-mcdonald}. The systematic forces on the colloidal particles are due to mutual interaction and the external electric field.

The structure of a system is characterized via the pair correlation function\cite{hansen-mcdonald}. This gives the probability distribution of separation between  particles in the system over equilibrium configurations. We construct analogous quantity to characterize spatial arrangements of particles. Since the system evolves in time, we need to generalize the pair correlation function to accommodate the temporal information. We do this by constructing equal time density correlation functions (ETDCF), namely, the  probability distribution of particle separations at a given time.  We calculate  structural ETDCF for the particles, irrespective of their mobilities, in the plane transverse to the field  to capture structural arrangement of particles in the transverse plane. Since  diffusivity involves squared displacements\cite{hansen-mcdonald}, we identify slow (S) and fast particles (F) from the distributions of squares of particle displacements in a given time interval. We  compute dynamic ETDCFs between S and F particles from their positions, treating them as different species.  The length scales of all the structural and dynamic ETDCFs are extracted from their envelope functions. We observe that the structural correlation lengths  grow with time as the lanes grow. Those of dynamic correlations for only the S particles grow, suggesting  that  the lateral structural arrangement is primarily dominated by the slow particles.

\section{Methods}
 We perform Brownian Dynamics simulations\cite{bd1,bd2,hansen-mcdonald} of binary charged colloid in water with an equal number (=5000) of positive and negative charges in a cubic simulation box of size $L=36.827\sigma$ in three dimensions with the periodic boundary conditions, as detailed in Ref. \cite{dutta3,lane5,lane10,lane13}.  The salient feature is that the equation of motion of a colloidal particle in the overdamped limit has been numerically solved. The systematic forces acting on a particle  are:  the viscous drag due to water medium of viscosity of 1.0 cP, the interaction forces given by the Derjaguin-Landau-Verwey-Overbeek (DLVO) potential, and the external electric field,  while the random force acting on the particles is given by Gaussian white noise with variance satisfying the fluctuation-dissipation \cite{dutta3,lane5,lane10,lane13}. We take $\tau_{\beta}$, time required to diffuse over one particle diameter, as unit of time and particle diameter $\sigma$ as unit length. We first equilibrate the system without electric field from random configurations. Then we switch on the electric field with strength, $f$ (in the unit of $\frac{k_{B}T}{\sigma}$, $k_{B}$ is the Boltzmann constant and $T=300 K$ the temperature) in $z-$ direction to drive the system away from equilibrium. Earlier studies\cite{dutta3,lane5,lane10} show lanes proliferated along $z-$direction and network like patterns in the transverse plane, consisting of like charges in steady state for $f(=300)$.  We use $f=300$ in our simulations and the field is kept on for $15\tau_{\beta}$ so that the system reaches steady state. We calculate different quantities at time $t_{w}$ after the field is turned on. The simulations are repeated over 10 different independent runs. 
 
 The ETDCF at time $t_{w}$ is defined as: 
 \begin{equation}
 g(r_{\bot},t_{w})=(1/N^{2})<\sum _{i=1}^{N}\sum _{j \neq i}\delta (r_{\bot} -|(\vec{ R_{j,\bot}}(t_{w})-\vec{R_{i,\bot}}(t_{w})|>    
 \end{equation}
  where $r_{\bot}$ separation  and  $\vec{R_{\bot}}$ the projection vector of the particle coordinate in the  plane ($x-y,\bot$) transverse to the applied field along z-direction. $<>$ implies average over independent sets of BD trajectories. One can note that this is a simple generalization of liquid pair correlation function in equilibrium \cite{hansen-mcdonald}.
  
We obtain structural ETDCF using the in-plane particle coordinates in Eq.1, irrespective of their mobilities. We construct mobility maps of mean squared displacements following coarse graining procedure explained in Refs. \cite{pc1,pc2} to show qualitatively the coexistence of particles with different mobilities. We  identify  a particle slow(S) in the system \cite{dutta3,donati} if its squared displacement is smaller  than the mean of the probability distribution of squared displacements in a given time interval. Similarly, a particle is identified as fast (F) if it has larger squared displacement than the mean. The particles with squared displacements beyond the $2\Sigma$ ($\Sigma$ the standard deviation of the distribution) limit have been considered for identifying S and F particles.We obtain dynamic ETDCFs by  computing ETDCF from in-plane coordinates of the S and F particles. The correlation lengths ($\xi$) of  structural ETDCFs and dynamic ETDCFs are extracted fitting of the associated envelopes ($U(r)$) to a form $U(r)\sim \exp(-r/\xi)$\cite{jc} which we call structural and dynamic correlation lengths respectively. We limit the fitting up to $r\approx 6\sigma$ for numerical accuracy.

\section{Results and discussions}

Fig. 1 shows  equilibrated snapshots of particles with coordinates projected onto transverse, XY plane  for a Brownian dynamics trajectory. Fig. 1(a) shows no significant structure at $t_{w}=0$. However, as soon as the field ($f=300$) is turned on, the particle arrangement undergoes changes. At small $t_{w}$($\approx 0.9\tau_{\beta}$),  tiny domains of particles of a given charge are observed [Fig. 1(b)]. At larger time $t_{w}=10\tau_{\beta}$ the particles of a given charge form network like patterns [Fig.1(c)]. Thus the charges undergo micro-phase separation with increasing time. Similar structural changes are observed for all other trajectories.

\begin{figure}[h]
	\begin{center}
		\includegraphics[angle=0,scale=0.72]{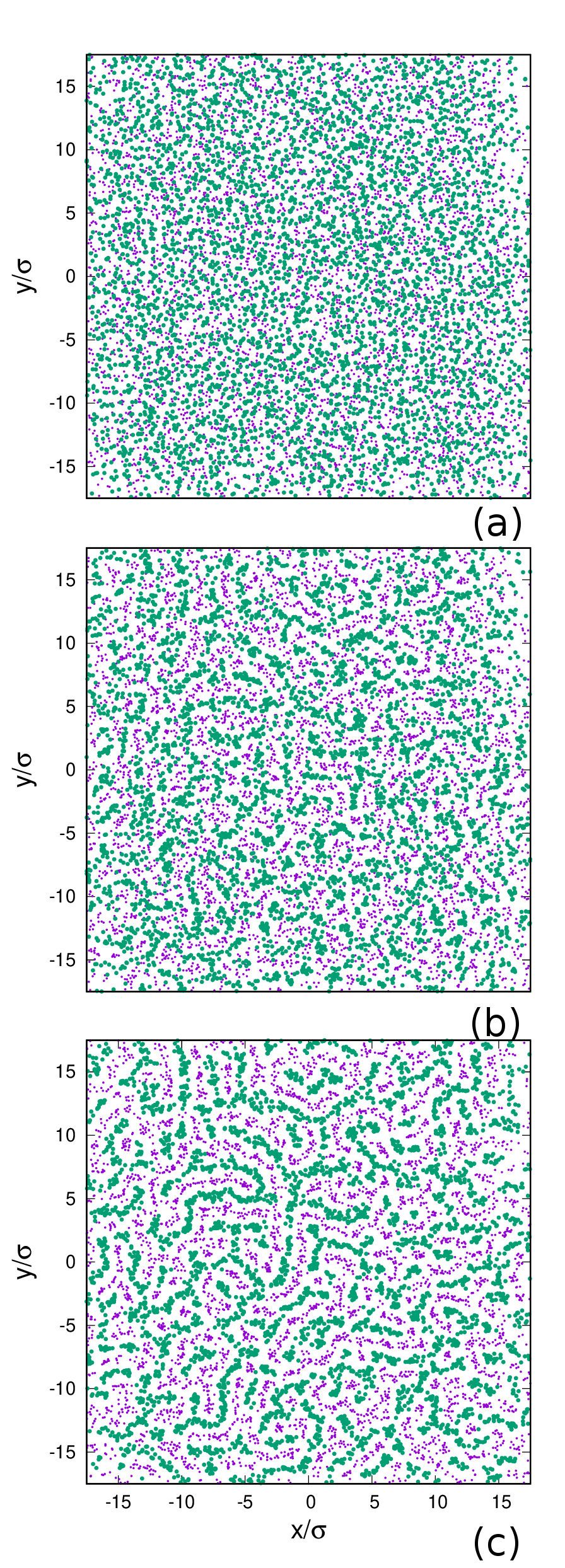}
		\caption{ (Colour Online) Particle configurations projected onto the transverse plane (x-y) for (a) $t_{w}=0$ (b) $t_{w}=0.5\tau_{\beta}$ (c) $t_{w}=10\tau_{\beta}$. Here $+ve$ and $-ve$ charges are shown in cyan and magenta dots respectively. }
	\end{center}
\end{figure}

The structural changes in the transverse plane take place concurrently with the growth of the lane order parameter $(\Phi)$, defined in Ref.\cite{lane5}, $\Delta \Phi(t_{w})(=\Phi(t_{w})-\Phi(0))$.  The order parameter  reaches steady values $(\Delta \Phi(t_{w})\approx 0.25)$ around $\tau_{S}\approx 1.2\tau_{\beta}$ in agreement to earlier report\cite{lane13}. The particle arrangement reaches steady pattern around the same time. Here, $t_{w}<\tau_{S}$ corresponds to transient behavior. On the other hand,  $t_{w}>\tau_{S}$ corresponds to the steady state conditions. 

{\bf Structural Correlations:}

\begin{figure}[h]
\begin{center}
\includegraphics[angle=0,scale=0.4]{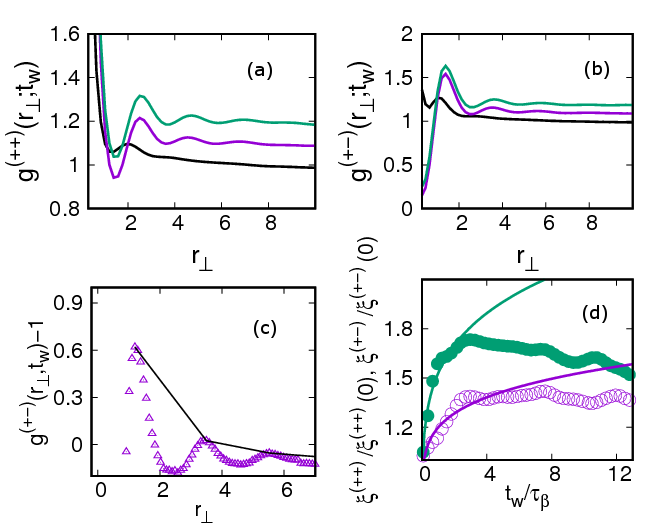}
\caption{ (Colour online) Dependences of (a) $g^{(++)} (r_{\bot};t_{w})$ and (b) $g^{(+-)} (r_{\bot};t_{w})$ with $r_{\bot}$ for $t_{w}=0.2 \tau_{\beta}$  (black line), $t_{w}=0.5 \tau_{\beta}$ (magenta, with vertical offset 0.1 unit) and $t_{w}=1.5 \tau_{\beta}$ (cyan, with vertical offset 0.2 unit). (c)  A typical example constructing envelope  $U^{(+-)} (r_{\bot};t_{w})$ over the peaks of $g^{(+-)} (r_{\bot};t_{w})-1$.  The solid line shows $U^{(+-)} (r_{\bot};t_{w})$. $U^{(++)}(r_{\bot};t_{w})$ is constructed similarly for other cases. (d) $\xi_{\bot}^{(++)}$ (magenta open circles) and $\xi_{\bot}^{(+-)}$ (cyan filled circles) as functions of $t_{w}$. The lines show $\sim t^{0.2}_{w}$ dependences in the transient condition. \label{fig71}}
\end{center}
\end{figure}

We now consider the structural ETDCFs irrespective of particle mobility. Because of symmetric mixture, both charges behave similarly and we focus on the $+ve$ charged particles. The structural ETDCFs between two $+ve$ charges is denoted by $g^{(++)} (r_{\bot};t_{w})$ which gives the probability that a pair of positively charged particles are separated by $ r_{\bot}$ in the transverse plane at a given time. This quantity has been computed using Eq.1 from the in-plane coordinates of the positive charged particles and averaged over all the Brownian dynamics trajectories. Similarly, the structural ETDCFs between two opposite charges is denoted by $g^{(+-)} (r_{\bot};t_{w})$ from the in-plane coordinates of the oppositely charged particles at a given instance and gives the probability of finding a pair of opposite charges at distance $ r_{\bot}$ in the transverse plane at that instance. Fig. \ref{fig71}(a) shows typical data for $g^{(++)} (r_{\bot};t_{w})$ and those for  $g^{(+-)} (r_{\bot};t_{w})$ are shown in Fig. \ref{fig71}(b).  Let us consider first the transient condition. We observe that  $g^{(++)} (r_{\bot};t_{w})$ has strong peak at $r_{\bot}\approx 0$, for all $t_{w}$ which we denote by $g^{(++)} (0,t_{w})$. At $t_{w} \approx 0.2\tau_{\beta}$, $g^{(++)} (r_{\bot};t_{w})$ develops a tiny first coordination shell at $r_{\bot}\approx 2 \sigma$. At higher $t_{w}(\approx 0.5\tau_{\beta})$, this peak gets sharper. At $t_{w}\approx 1.5\tau_{\beta}$ corresponding to the steady state, the peak broadens and shifts to higher values in $r_{\bot}$ and more peaks appear. This corresponds to the network like patterns of micro-phase separated domains in the transverse plane. The alignment into lanes in z-direction gives rise to the peak in  $g^{(++)} (0,t_{w})$. This quantity can be taken as a measure of lane formation. We find that  $g^{(++)} (0,t_{w})$ increases  with increasing $t_{w}$ in transient condition and finally saturates to a steady value (~8.0) at the same time as the lane order parameter\cite{lane13}. 

Fig. \ref{fig71}(b) shows that $g^{(+-)} (r_{\bot};t_{w})$ has a minimum at $r_{\bot}\approx 0$ and a peak at $r_{\bot}\approx 1.1\sigma$ for $t_{w}=0.2\tau_{\beta}$. At $t_{w}\approx 0.5\tau_{\beta}$, the peak shifts to $r_{\bot}\approx 1.3\sigma$. This feature remain similar in the steady state, $t_{w}=1.5\tau_{\beta}$. The minimum $g^{(+-)} (r_{\bot};t_{w})$ at $r_{\bot}\approx 0$, denotes that lanes are formed by expulsion of negative charges from the domain of the positive charges. The appearance of more peaks with increasing time denotes  formation of domains  rich in one charge via micro-phase separation of the charges in the transverse plane, consistent with the particle snapshots. 

 In order to quantify the structural correlation length, we construct an envelope function $U^{(+\pm)}(r_{\bot};t_{w})$ as follows: We consider only the peaks in $g^{(+\pm)} (r_{\bot};t_{w})-1$.  The subtraction is done to ensure that we consider only the correlated part \cite{hansen-mcdonald} of the structural ETDCF. A smooth function is constructed through the peak points which is the envelope function. A typical envelope function is given in Fig. \ref{fig71}(c). We find that $U^{(+\pm)}(r_{\bot};t_{w})$ decay exponentially with $r_{\bot}$. We obtain the structural correlation length at time $t_{w}$, $\xi_{\bot}^{(+\pm)}(t_{w})$ by fitting $U^{(+\pm)}(r_{\bot};t_{w})$ to a dependence $\sim \exp(-r_{\bot}/\xi^{+\pm}_{\bot}(t_{w}))$ \cite{jc}.

 We observe that $\xi_{\bot}^{(++)}(0) \approx \xi_{\bot}^{(+-)}(0) \approx \sigma$ as in the equilibrium system. We show in Fig. \ref{fig71}(d) the dependences of $\xi_{\bot}^{(++)}$ and $\xi_{\bot}^{(+-)}$ on $t_{w}$. Both $\xi_{\bot}^{(++)}$ and $\xi_{\bot}^{(+-)}$ remain finite. $\xi_{\bot}^{(++)}$ increases in transient condition till $t_{w}~\tau_{S}$  and reaches a steady value $1.4 \xi_{\bot}^{(++)}(0)$ in the steady state for $t_{w}>\tau_{S}$. $\xi_{\bot}^{(+-)}$, on the other hand, increases till $t_{w}\simeq\tau_{S}$  and then decreases slowly for $t_{w}>\tau_{S}$ with a maximum ($1.6 \xi_{\bot}^{(+-)}(0)$) $t_{w}\approx \tau_{S}$. 
 
 The time dependence of the structural length scales indicate arrangement of particles evolving with time. Fig. \ref{fig71}(d) shows that both $\xi_{\bot}^{(++)}$ and $\xi_{\bot}^{(+-)}$ follow algebraic dependence ($\sim t_{w}^{\alpha}$) with $\alpha\approx 0.2$ in the transient conditions. The  increase $\xi_{\bot}^{(++)}$ in  transient condition is in agreement to tendency of the positive charge particles to form  domains  in the transverse plane. This continues till the domains proliferate the plane when this correlation length saturates. The micro-phase separation of positive particles proceed by displacing the negative particles from the mixed domains via correlated motions over a long distances which is reflected in increase in $\xi_{\bot}^{(+-)}$ in the transient condition. The decrease in $\xi^{(+-)}$ for $t_{w}>>\tau_{S}$ indicates that once positive particles enrich the in-plane domains proliferating the transverse plain, the negative charges stay away from the domains completely. 

\begin{figure}[h]
	\begin{center}
		\includegraphics[angle=0,scale=0.2]{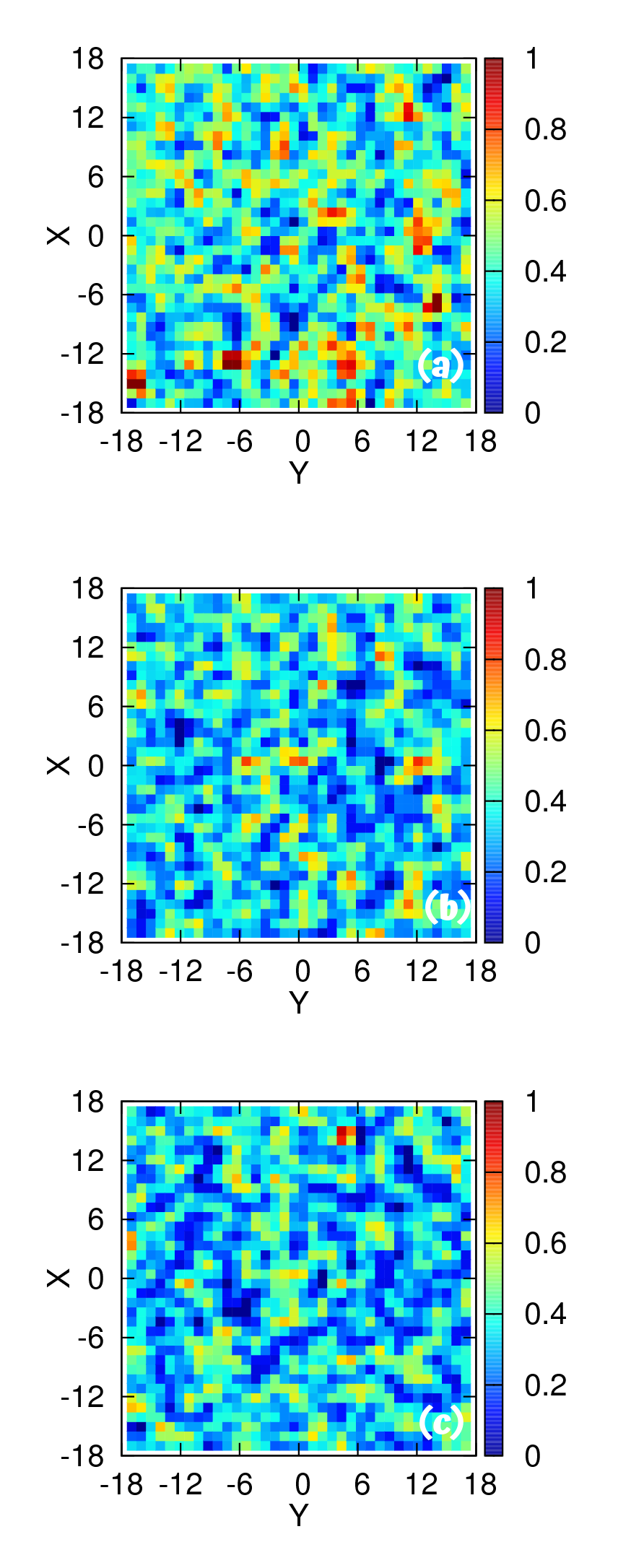}
		\caption{ (Colour online) Coarse-grained spatial maps of $\delta r^{2}_{\bot}$ for the $+ve$ charges at (a) $t_{w}=0.2 \tau_{\beta}$ (b)$t_{w}=\tau_{\beta}$ and (c) $t_{w}=5 \tau_{\beta}$ for an interval $\delta t=0.75\tau_{\beta}$ at coarse-grain scale $\sim \sigma$.}
	\end{center}
\end{figure}

{\bf Dynamic correlations:} 

 In normal systems the mobilities of the particles show homogeneity with  well defined diffusion coefficient\cite{hansen-mcdonald}. However, lane forming colloids have different diffusivities, as shown earlier\cite{lane10,lane13}. We make an attempt to see the spatial distribution of particle with different diffusivities. Since diffusion coefficient is given in terms of mean squared displacements,  we identify the particle mobility in terms of square of the particle displacements in a given time interval. We consider two configurations at time $t_{w}$ and $t_{w}+\delta t$ and compute the square of the particle displacements in the transverse plane, $\delta r^{2}_{\bot}$.  We construct coarse-grained spatial map of squared displacements in the transverse plane in the given time interval over a given Brownian trajectory as follows. The $\bot$ plane is divided into  $40~X~40$ spatial zones for coarse graining with reasonable statistics. We locate a particle in the $\bot$ plane  at $t_{w}$ and compute the displacement between time $t_{w}$ and $t_{w}+\delta t$, $\delta r_{i}^{2}(t_{w},\delta t)=|\vec{R_{i,\bot}}(t_{w})-\vec{R_{i,\bot}}(t_{w}+\delta t)|$ in the transverse plane. We average the values of the squared displacements over  the particles of +ve charge  in each grid to obtain the  mobility map at time $t_{w}$ over the spatial grid.  We fix $\delta t(=0.75\tau_{\beta})$ such that the changes in $\delta r^{2}_{\bot}$ are appreciable. We show maps for three different $t_{w}$. The map shows the organization of particles  with  different mobilities. The slow particles show string like patterns proliferating in the system with increasing $t_{w}$. The negative particles proliferate along $z$-direction in lanes in the blue patches (data not shown).

 \begin{figure}[h]
	\begin{center}
		\includegraphics[angle=0,scale=0.5]{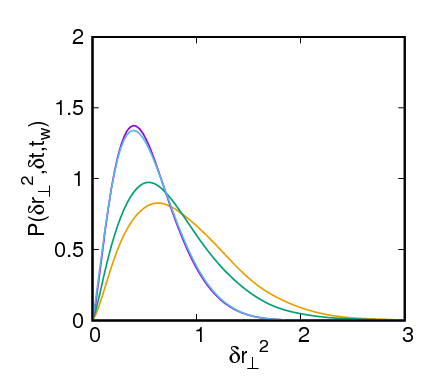}
		\caption{(Colour online) (a) $P(\delta r^{2}_{\bot}; \delta t,t_{w})$ vs $r^{2}_{\bot}$ for $t_{w}=0.3 \tau_{\beta}$ (yellow), $t_{w}=0.5\tau_{\beta}$ (green), $t_{w}= 1.1\tau_{\beta}$ (blue) and $t_{w}= 10\tau_{\beta}$ (magenta) \label{fig72}}
	\end{center}
\end{figure}   
 
The mobility maps given in Fig.3 are rather qualitative.  We observe that the maps are different for each of the trajectories. In order to quantify the heterogeneity in mobilities over different Brownian trajectories, we calculate the probability distribution of the squared displacements of the  positive charged particles $P(\delta r^{2}_{\bot}; \delta t,t_{w})$, obtained by binning the square of the particle displacements (without coarse graining) in a given interval. This quantity is computed using all the simulated trajectories. We show In Fig. \ref{fig72} the evolution of $P(\delta r^{2}_{\bot}; \delta t,t_{w})$ as a function of $\delta r^{2}_{\bot}$ for different $t_{w}$. We choose $t_{w}$ values as follows: a couple of cases before the steady state sets in, another close to the steady state and the other one deep into the steady state.  We observe that for all $t_{w}$, the distribution is asymmetric in $\delta r^{2}_{\bot}$ with respect to the peak value. With increasing $t_{w}$, the peak in $P(\delta r^{2}_{\bot}; \delta t,t_{w})$ shifts to lower values of $\delta r^{2}_{\bot}$, indicating slowing down in the system. The negatively charged particles show similar mobility distributions (data not shown).

 We calculate the mean $\mu (t_{w})$ and standard deviation $\Sigma(t_{w})$ of the distribution. We tag the particles as slow (S) and fast (F) as follows: We consider particles excluding $2\Sigma(t_{w})$ region in mobility about the mean to identify fast and slow particles. A particle of a particular species is tagged  "slow" (S) if it experiences squared displacement  $\delta r^{2}_{\bot}<\mu (t_{w})-\Sigma(t_{w})$. Similarly, we tag the particles with $\delta r^{2}_{\bot}\geq \mu (t_{w})+\Sigma(t_{w})$ as "fast" (F).We identify $N^{(\pm)}_{S}(t_{w})$ no of S particles of $+ve$ and $-ve$ charges respectively in window of $t_{w}$ and $t_{w}+ \delta t$. Similarly, we count $N^{(\pm)}_{F}(t_{w})$ F particles in the same time window. A choice of higher $\Sigma(t_{w})$  does not impact much the framework but the quality of data get too poor to have reasonable statistics due to decrease in $N^{(\pm)}_{F}(t_{w})$ and $N^{(\pm)}_{S}(t_{w})$.

\begin{figure}[h]
\begin{center}
\includegraphics[angle=0,scale=0.9]{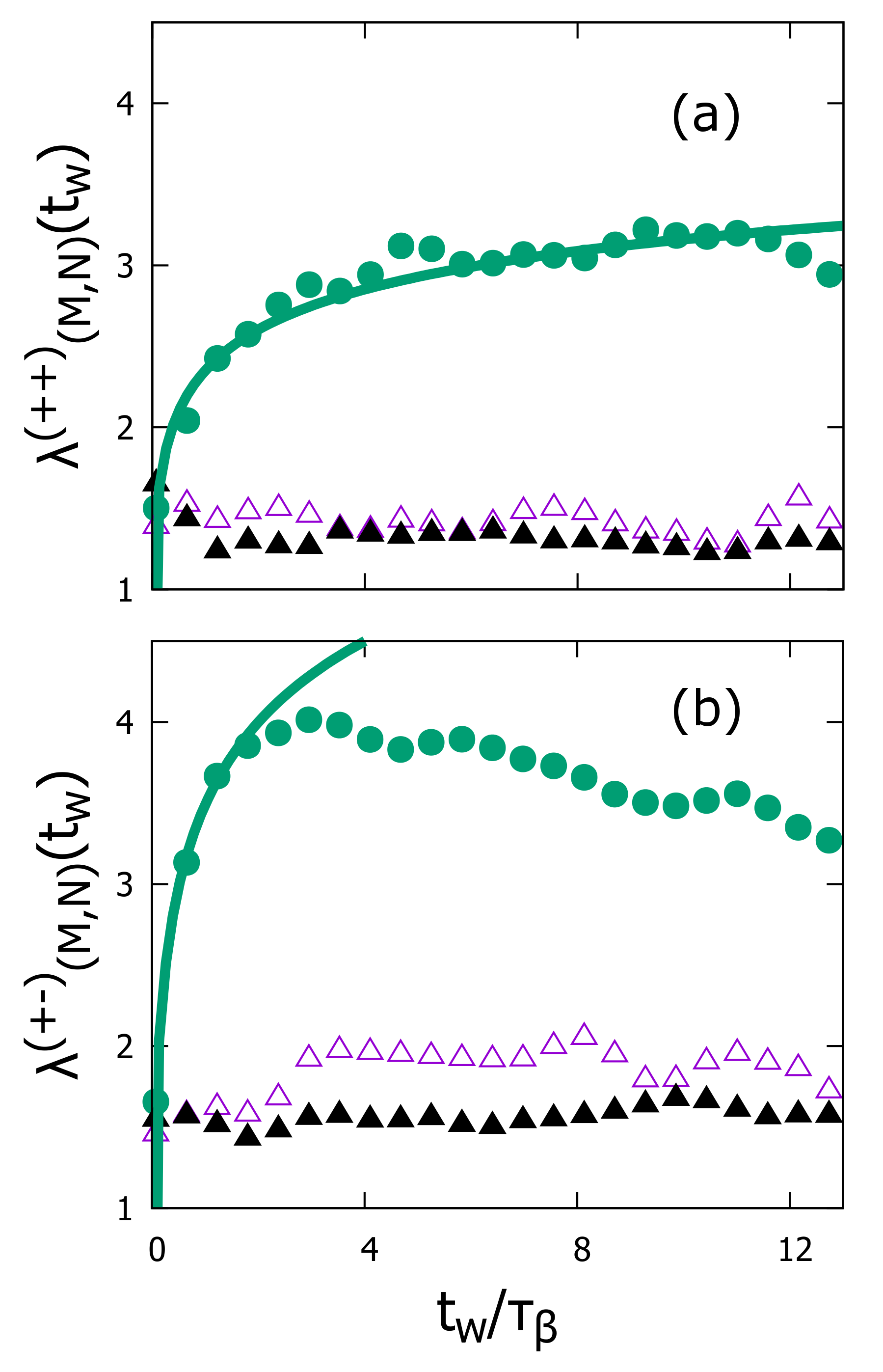}
\caption{ (Colour Online) (a) $ \lambda^{(++)}_{(M,N)}$ and (b) $\lambda^{(+-)}_{(M,N)}$ as functions of $t_{w}$.  In both panels the symbols are as follows: $M=S$ and $N=S$ (cyan filled circles), $M=F$ and $N=S$ (magenta open triangles), $M=F$ and $N=F$ (black filled triangles). Dashed lines show $\sim t_{w}^{\alpha}$ dependence with $\alpha\approx 0.16$  in (a) and $\alpha\approx 0.2$ in (b). \label{fig73}}
\end{center}
\end{figure}

\begin{figure}[h]
\begin{center}
\includegraphics[angle=0,scale=0.65]{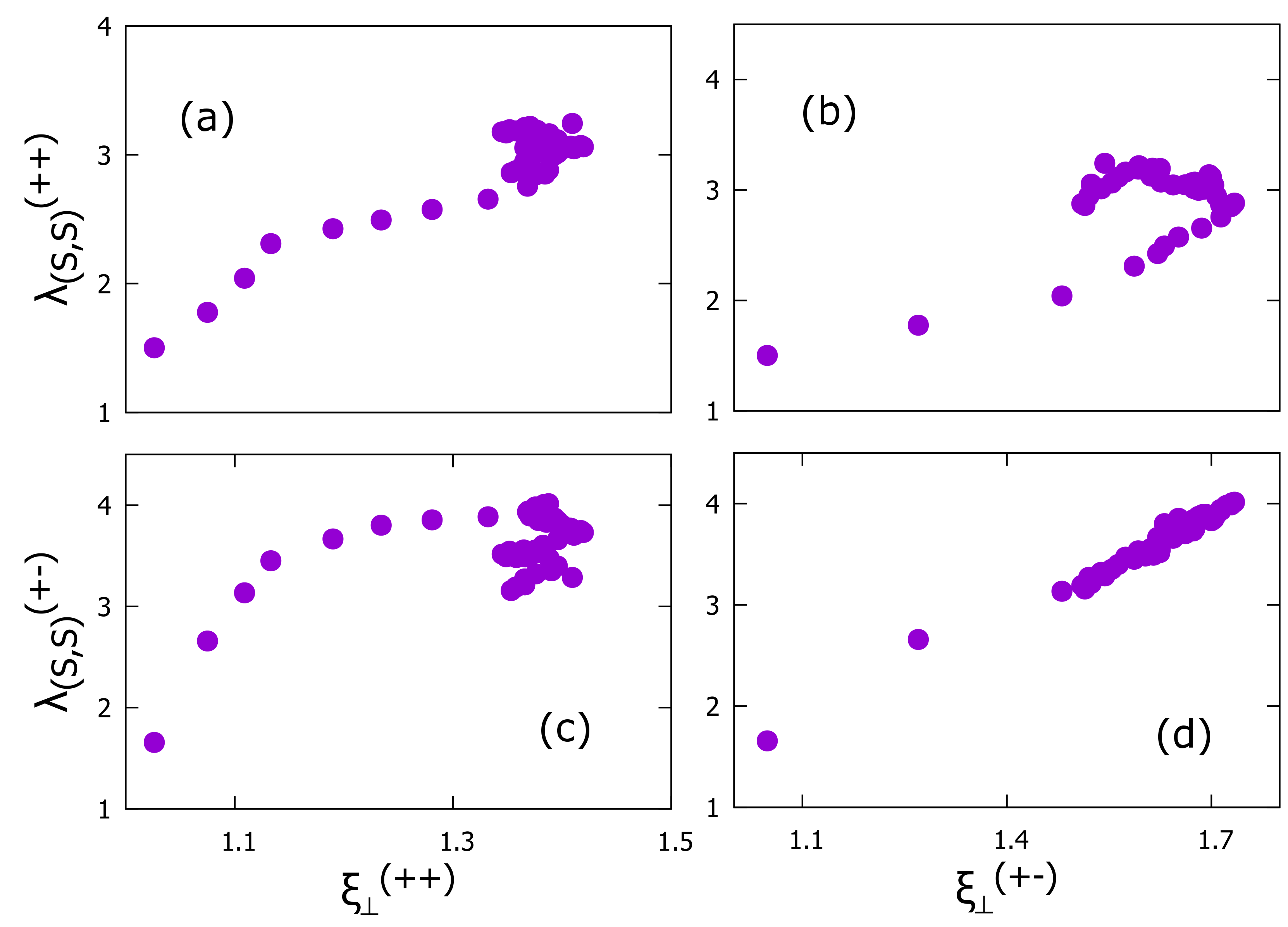}
\caption{Correlation plots: (a)$ \lambda^{(++)}_{(S,S)}$ vs $\xi_{(\bot)}^{(++)}$ (b)$ \lambda^{(++)}_{(S,S)}$ vs $\xi_{(\bot)}^{(+-)}$ (c)$ \lambda^{(+-)}_{(S,S)}$ vs $\xi_{(\bot)}^{(++)}$ (d)$ \lambda^{(+-)}_{(S,S)}$ vs $\xi_{(\bot)}^{(+-)}$ \label{fig74}}
\end{center}
\end{figure}

We now construct the dynamic ETDCFs of the S and F species from their in-plane coordinates at $t_{w}$ using Eq.1 to characterize their spatial arrangement at a given instance, considering all the simulated trajectories. There are now total six possibilities of the dynamic ETDCFs: for each charge there are both types of paricles and also their cross correlations. We denote them by $g^{(+\pm)}_{M,N}(t_{w},r_{\bot})$ where $M=F,S$ and $N=F,S$. The interpretation of $g^{(++)}_{S,S}(t_{w},r)$ is  the probability of finding  a pair of S positive particles at a separation $r_{\bot}$  in the transverse plain at time $t_{w}$. Similar interpretation holds for the other dynamic ETDCFs. Thus, the dynamic ETDCFs give the structural order at a given time instance among the particles of different mobilities. 

As in the structural ETDCF we fit envelope function through the peaks of the dynamic ETDCFs in order to extract the dynamic correlation lengths. The envelopes $U_{(M,N)}(r_{\bot},t_{w})$ show exponential decay in $r_{\bot}$ and are fitted by $\exp(-r_{\bot}/\lambda_{(M,N)}^{(+\pm)}(t_{w}))$. Here, $\lambda_{(M,N)}^{(+\pm)}(t_{w})$ are the dynamical correlation lengths among the species of different mobilities. We show six possible dynamic correlation lengths in Fig. \ref{fig73}(a) and (b) as functions of $t_{w}$. Out of these six,  $\lambda_{(S,S)}^{(++)}$ and $\lambda_{(S,S)}^{(+-)}$ increase with $t_{w}$ in the transient condition,  while the others are independent of time. Both $\lambda_{(S,S)}^{(++)}$ and $\lambda_{(S,S)}^{(+-)}$ become quite large, extending upto 3-4 particle diameters. We observe $\lambda_{(S,S)}^{(++)}(t_{w})\sim t_{w}^{0.16}$ for the entire time range. However, $\lambda_{(S,S)}^{(+-)}(t_{w})\sim t_{w}^{0.2}$ in the transient regime and then there is decrease in this length. The increase in these two length scales in transient condition suggests time dependent rearrangement  of S positive particles in the transverse plane, consistent with Fig.3. 

 It is interesting to check if the clustering of S +ve particles is coupled to the micro-phase separation of the positive charges .This is best reflected in the correlation diagram of different correlation lengths. Fig. \ref{fig74}  show such correlation diagrams by eliminating time between the structural correlation $\xi^{(++)}$ and  $\xi^{(+-)}$ and the dynamic correlation lengths $\lambda_{(S,S)}^{(+\pm)}$ . Fig. \ref{fig74} (a) shows that $\lambda^{(++)}_{(S,S)}$ increases linearly with $\xi_{\bot}^{(++)}$ in the transient conditions (upto $\xi^{(++)}=1.4$). After the steady state is reached,  these correlation lengths do not change much. The linear increase in both length scales suggests that the development of the structural order in the micro-phase separated domains  is governed by the slow species only.  $\lambda^{(++)}_{(S,S)}$ does not change after the structural arrangement stabilizes. Similarly, $\lambda^{(++)}_{(S,S)}$ increases with $\xi_{\bot}^{(+-)}$ upto the steady ($\xi^{(+-)}/\xi^{(+-)}(0)=1.6$) [Fig. \ref{fig74}(b)]. In the steady state  $\xi_{\bot}^{(+-)}$ decreases, leading to the upper part of the correlation curve.  This indicates that the micro-phase separation of the positive particles in the transverse plane involves  arrangement of the slow positive particles over a large length scale, while the negatively charged slow particles are displaced. 
 
 $\lambda^{(+-)}_{(S,S)}$ increases linearly  with $\xi_{\bot}^{(++)}$ initially after which this length-scale shows saturation till the steady-state is reached and in the steady state condition, the dynamical correlation length decreases, while the structural correlation length does not change [Fig. \ref{fig74} (c)]. This data suggest that the expulsion of the negatively charged slow particles goes on, although the domains structurally do not evolve much. This reflects the coarsening of the structural domains and those of S particles in the transverse plane concurrently. On the other hand, $\lambda^{(+-)}_{(S,S)}$ increases with $\xi_{\bot}^{(+-)}$ [Fig. \ref{fig74} (d)]which is also consistent with  concurrent coarsening of the domains. 
 
 Our data suggest that the slow S particles tend to cluster together in the background of the fast F particles.  A slow particle prefers to be in the vicinity of another slow particle leading to mobility resolved phase separation\cite{tanaka,dutta3,pedx} known for systems out of equilibrium. The slowing down of the system as the steady state is approached further supports that the nucleation of the structure proceeds via the slow particles. The late time structural morphology [Fig.1(c)] the in-plane structure exhibited by the slow particles in the steady state [Fig.3] have overall similarity which qualitatively supports the concurrent growth of domains of slow particles and structural domains of a given charge in the system. Physically, faster particles can easily perturb the ensuing structure in the system, while slower particles have much longer stabilization time in each other's vicinity that helps the overall structure to grow. Thus, the kinetics of changes in structural morphology can be understood from those of slow particles in the system. Late time structural morphology is quite similar to the morphology of patterns in various systems out of equilibrium \cite{new1} where a similar scenario is expected to hold.
 
 \section{Conclusion}
 
To summarize, our data show that the structural arrangement in the transverse plane in lane-forming colloid proceeds via clustering of the slow particles in the background of fast-moving particles. We show the existence of dynamic-correlation with well defined dynamical correlation length in the system. The colloidal system has a unique advantage that their motions can be followed by video microscopy\cite{vmic}. One can, therefore, directly construct the structural and dynamic correlation functions. It will be interesting to study in future the nucleation of slow particles which could be pertinent to characterize the kinetics of pattern formation in terms of dynamic correlations in far from equilibrium conditions. 
 
We thank C. Dasgupta and M. Dijkstra for insightful discussions. This research was supported in part by the International Centre for Theoretical Sciences (ICTS) during a visit for the program - Indian Statistical Physics Community Meeting (Code: ICTS/ispcm2019/02). There are no conflicts to declare.

\end{document}